\documentclass[pra,twocolumn]{revtex4-1}
\usepackage{amsmath}
\usepackage{amsfonts}
\usepackage{amsbsy}
\usepackage{xcolor}
\usepackage{soul}
\usepackage{graphicx}

\newcommand{\sinc}{\mathrm{sinc}}

\newcommand{\spvec}[1]{\ensuremath{\mathbf{#1}}}
\newcommand{\unitvec}[1]{\ensuremath{\mathbf{\hat{#1}}}}

\newcommand{\colvec}[1]{\ensuremath{\mathrm{#1}}}

\newcommand{\commentout}[1]{{}}

\newcommand{\rv}{\spvec{r}}

\newcommand{\beq}{\begin{equation}}
\newcommand{\eeq}{\end{equation}}
\newcommand{\rmE}{\mathrm{E}}
\newcommand{\rmM}{\mathrm{M}}
\newcommand{\sjE}{\rmE}
\newcommand{\sjM}{\rmM}
\newcommand{\sjI}{\mathrm{I}}
\newcommand{\sjS}{\mathrm{S}}

\newcommand{\sjT}{\mathrm{T}}

\begin{document}
\title{Metamaterial transparency induced by cooperative
  electromagnetic interactions}
\author{Stewart D. Jenkins}
\author{Janne Ruostekoski}
\affiliation{School of Mathematics and Centre for Photonic
  Metamaterials, University of Southampton,
  Southampton SO17 1BJ, United Kingdom}
\begin{abstract}
  We propose a cooperative asymmetry-induced transparency, CAIT, formed
  by collective excitations in metamaterial arrays of discrete resonators.
  CAIT can display a sharp transmission resonance even when the
  constituent resonators individually exhibit broad resonances.
  We further show how dynamically reconfiguring the metamaterial
  allows one to actively control the transparency. While
  reminiscent of electromagnetically induced transparency, which can be
  described by independent emitters, CAIT relies on a cooperative
  response resulting from strong radiative couplings between the
  resonators. 
\end{abstract}
\date{\today}
\maketitle

Electromagnetically induced transparency (EIT), a result of
destructive interference between different excitation paths, causes an
otherwise opaque collection of electromagnetic (EM) emitters to become transparent over
a range of frequencies.
In atomic gases, interference between atomic
level transitions prevents the excitation of a transition that
scatters incident light
\cite{BollerEtAlPRL1991,HarrisPhysToday1997,FleischhauerEtAlRMP2005}.
This interference abruptly alters the dispersion relation for frequencies in
the transparency window, providing a
mechanism to slow \cite{HauEtAlNature1999} or even
stop light for later retrieval
\cite{LiuEtAlNature2001,LukinImamogluNature2001}.
Slow and stopped light pulses have lead to applications in sensitive
magnetometry
\cite{FleischhauerScullyPRA1994,KatsoprinakisEtAlPRL2006,BudkerRamalisNatPhys2007,YudinEtAlPRA2010},
all-optical switching \cite{BahramiEtAlJMO2010}, and quantum
memories
\cite{FleischhauerLukinPRL2000,ChanelierePhotStorNAT2005,ChoiEtAlNature2008,JenkinsEtAlJPB2012A}.

Several theoretical proposals
\cite{ZhangEtAlPRL2008,TassinEtAlPRL2009,LuLiuMauPRA2012,VerslegersEtAlPRL2012} and experimental realizations 
\cite{PapasimakisEtAlPRL2008, PapasimakisEtAlAPL2009,
LiuEtAlNatMat2009,LiuEtAlNanoLett2010,ZhangEtAlAPL2010,ZhangEtAlOpEx2010,ZhangEtAlAPL2011,KurterEtAlPRL2011,TassinEtAl2012} have transferred the idea of EIT  in
independently scattering atoms to metamaterial arrays
of circuit elements.
In these artificially structured materials, the unit-cell resonators
(meta-molecules) play a role analogous to atoms in conventional EIT.
A transmission resonance forms via coupling between
two modes of plasmonic excitations in independently scattering
metamolecules: a bright mode that strongly radiates and a dark mode
with a narrower radiative linewidth.
Broad radiative linewidths of nanofabricated circuit elements,
however, can severely limit the quality of EIT-like transmission
resonances in independently scattering metamolecules.

Recent studies
\cite{JenkinsLongPRB,JenkinsLineWidthNJP,FedotovEtAlPRL2010,KAO10,AdamoEtAlPRL2012}
have shown that, rather than independently, certain systems of
closely-spaced resonators respond cooperatively to an incident field.
In particular, interactions between resonators that are mediated by
scattered EM fields result in collective modes of resonator
excitations \cite{JenkinsLongPRB}, several of which have significantly
narrowed radiative linewidths.

In this Letter, we show how to exploit such collective modes to
realize a cooperative transmission resonance.
We propose a cooperative asymmetry-induced transparency (CAIT) in
metamaterials.
Unlike transmission resonances based on independent scatterers, the
bright and dark modes in CAIT are collective.
Specifically, the dark mode possesses a cooperatively narrowed
resonance linewidth.
This narrowing leads to a sharp resonance of high
transmission, even though the resonators forming the metamaterial
would individually, in isolation, exhibit broad
resonances. 

The transmission resonance is sensitive to the size of the system and
the specific resonator configuration.
Limited only by intrinsic nonradiative losses, the transmission
resonance and group delay of a transmitted pulse can become
progressively narrower and longer, respectively, with increasing size
of a two-dimensional (2D) metamaterial array.
In a $205 \times 205$ array, for example, we estimate the resonance
width (pulse delay) to be approximately $\Gamma/1000$ ($1600/\Gamma$),
where $\Gamma$ is the linewidth of a single isolated resonator.
Furthermore, changing relative positions of the resonators alters the
EM mediated interactions between them, and hence the cooperative
material response.
We show that using reconfigurable metamaterials
\cite{PryceEtAlNanLett2010,OuEtAlNatNan2013}, in which one can dynamically shift the
layout of the metamolecules, allows one to actively control the
transparency.

\begin{figure}
  \centering
  \includegraphics{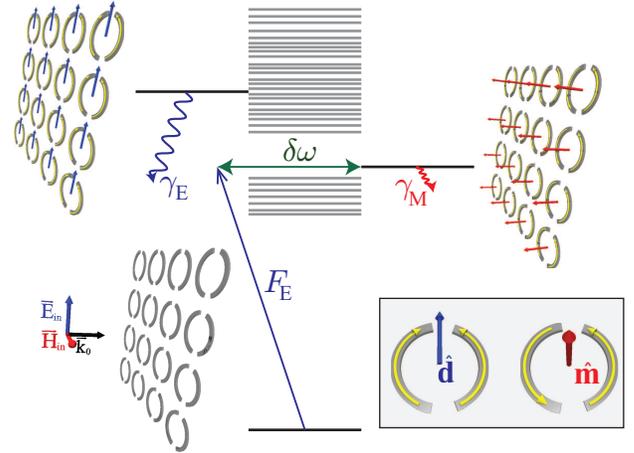}
  \caption{(Color online) A schematic illustration of CAIT in an array of ASRs. The
    inset shows symmetric (antisymmetric) currents in the ASR
    meta-atoms producing  electric (magnetic) dipoles along
    $\unitvec{d}$ ($\unitvec{m}$). The PME (PMM) mode is illustrated in the upper left
    (upper right), and has a decay rate $\gamma_{\sjE}$ ($\gamma_{\sjM}$).
    An incident wave couples an unexcited (lower) array to the PME mode with
    strength $F_{\sjE}$, while
    asymmetry $\delta\omega$ induces a coupling between the PME
    and PMM modes.
    The grey lines represent other collective modes.
  }
  \label{fig:LevelStructure}
\end{figure}

To illustrate CAIT,  we consider a 2D array of asymmetric split rings
(ASRs)~\cite{FedotovEtAlPRL2007,FedotovEtAlPRL2010}, consisting of
pairs of concentric circular arcs (Fig.~\ref{fig:LevelStructure}). The
setup is closely related to recent transmission resonance
experiments~\cite{FedotovEtAlPRL2010}.  
In each ASR, currents can flow symmetrically, producing a net electric
dipole along the direction $\unitvec{d}$, or antisymmetrically,
producing a net magnetic dipole along $\unitvec{m}$.
CAIT forms from the coupling between two
phase-coherent collective modes of ASR excitations that are phase-matched with an incident EM plane wave
propagating perpendicular to the array.
The incident field drives the phase-matched electric (PME) mode,
dominated by all electric dipoles oscillating in phase,
while the
phase-matched magnetic (PMM) mode, consisting almost entirely of
magnetic dipoles perpendicular to the array,
does not directly couple to the incident field.
In an array with subwavelength lattice spacing,
the radiative linewidth of the PMM mode $\gamma_{\sjM}$
narrows with the system size.
Because it radiates only
weakly~\cite{JenkinsLineWidthNJP},
the PMM mode can be used as a collective dark-mode in CAIT.
For
example, in a 33$\times$33 array of split rings separated by half a wavelength, cooperative interactions reduce $\gamma_{\sjM}$ fifty fold
\cite{JenkinsLongPRB}.
An excited PME mode, on the other hand, radiates with rate $\gamma_{\sjE}\gg \gamma_{\sjM}$,
scattering the field into the forward and backward directions and
reflecting the incident field.
We will show how a transmission resonance with an active control forms via an interference
that permits excitation of the cooperatively narrowed PMM mode at the expense of the PME mode.

We consider a 2D square lattice of identical ASRs in the $z = 0$ plane
with subwavelength lattice-spacing $a$ and lattice vectors
$\spvec{a}_1 = a\unitvec{e}_x$ and $\spvec{a}_2 = a\unitvec{e}_y$.
The ASR electric (magnetic) dipoles -- produced by symmetric
(antisymmetric) current oscillations -- are oriented along
$\unitvec{d}= \unitvec{e}_y$ ($\unitvec{m} = \unitvec{e}_z$); Fig.~\ref{fig:LevelStructure}. 
Each ASR, labeled by index $\ell$ ($\ell =1 \ldots N$), comprises two meta-atoms (circular arcs).
A meta-atom, labeled by index $j$ ($j = 1 \ldots 2N$),
behaves as a radiatively damped LC circuit which is driven by the incident field and the
fields emitted by all other meta-atoms in the system
\cite{JenkinsLongPRB}.
We describe the current flow in meta-atom $j$ by a slowly varying complex amplitude
$b_j$. (See Appendix~\ref{sec:descr-asymm-split} for technical details.)
The meta-atom resonance frequencies are centered on $\omega_0$.
Owing to an asymmetry in arc lengths, the resonance frequencies of the right ($j =
2\ell$) and left ($j = 2\ell - 1$)
meta-atoms in each ASR are shifted by $\delta\omega$ and $-\delta\omega$,
respectively.
The oscillating electric and magnetic dipoles of each meta-atom
radiate at respective rates
$\Gamma_{\sjE}$ and  $\Gamma_{\sjM}$.

To better understand how a collection of ASRs behaves in concert, we
first examine a single, isolated ASR of two interacting arcs.
The dynamics of an ASR $\ell$ can be described
by the amplitudes of symmetric, $c_{\ell,+}$, and antisymmetric, $c_{\ell,-}$,
current oscillations, which are given in terms of the meta-atom
variables as $c_{\ell,\pm} =  \left(b_{2\ell} \pm b_{2\ell-1}\right) /
\sqrt{2}$.
These oscillations are eigenmodes of a single
symmetric split ring (SSR) ($\delta\omega=0$) with the radiative decay rates
$\gamma_+ \approx 2\Gamma_\sjE$ and $\gamma_{-}
\approx 2\Gamma_\sjM$ and resonance frequencies $\omega_0\pm \delta$.

In a single ASR the asymmetry shifts the resonance frequencies of the left and right arcs.
As a result, the symmetric and antisymmetric oscillations are no longer eigenmodes of a single ASR, and the evolution of those oscillations becomes coupled
\begin{equation}
  \label{eq:c_dynamics}
  \dot{c}_{\ell,\pm} = \left(-\gamma_{\pm}/2 \mp i\delta\right) c_{\ell\pm} -
  i\delta\omega c_{\ell,\mp} + F_{\ell,\pm} \textrm{ ,}
\end{equation}
where $F_{\ell,\pm}$ represents the external driving.
An EIT-like resonance of independently scattering ASRs
requires that $\gamma_\mp \ll \gamma_\pm$.
This would allow the dark mode (with lower emission rate) to be highly
excited so that the coupling $\delta\omega$ to the bright mode (with
higher emission rate) destructively interferes with
driving of the bright mode by the incident field.
In most experimental situations involving ASRs
\cite{FedotovEtAlPRL2007,FedotovEtAlPRL2010}, however, $\gamma_+$ and
$\gamma_-$ are comparable.
An array of independently scattering ASRs therefore cannot exhibit an
EIT-like transmission resonance.

The situation differs, however, in a metamaterial array of several ASRs that interact
via scattered EM fields.
As a result of interactions, the system possesses collective modes of oscillation
extended over the metamaterial.
To show how CAIT can emerge from these collective modes,
we construct an approximate phenomenological model from the PME and
PMM modes, the two
collective modes that are phase matched with the incident field.
We use this model to analytically calculate the steady-state
reflectance and transmittance.
The mode properties, the accuracy of the phenomenological model,
and the role of other collective modes in the metamaterial's EM response are numerically determined using the formalism
introduced in Ref.~\cite{JenkinsLongPRB}. 
These calculations fully incorporate all
dependent scattering processes~\cite{vantiggelen90, MoriceEtAlPRA1995,
RuostekoskiJavanainenPRA1997L,
RuostekoskiJavanainenPRA1997, JavanainenEtAlPRA1999,optlattice,
JenkinsLongPRB} between the resonators to all orders.
Applying the formalism to a 2D array of ASRs
\cite{JenkinsLineWidthNJP} yielded a narrowing of collective
linewidths with system size that agreed extremely well with
experimental measurements of transmission resonances~\cite{FedotovEtAlPRL2010}.

In the analysis, we approximate the incident EM field by a monochromatic plane wave and write
the positive frequency component of the electric field amplitude
$
\mathbf{E}^+_{\mathrm{in}} (\spvec{r},t) = \mathcal{E} \,\unitvec{e}_y e^{i\spvec{k}_{\mathrm{in}}\cdot \spvec{r}-i \Omega t}
$,
where $\unitvec{e}_y$ and $\spvec{k}_{\mathrm{in}}=k \unitvec{e}_z$
($k=\Omega/c$) denote the polarization and wavevector,
respectively.
The collective dynamics of the full metamaterial system,
described by meta-atom
variables $\colvec{b} \equiv (b_1, b_2, \ldots, b_{2N-1}, b_{2N})^T$,
is governed by \cite{JenkinsLongPRB,JenkinsLineWidthNJP}
\begin{equation}
  \label{eq:ColDyn}
  \dot{\colvec{b}} = \mathcal{C}\colvec{b} + \colvec{F}(t),\quad \mathcal{C}=\mathcal{C}_{\mathrm{SSR}}
    -i\delta\omega\mathcal{A} \, \textrm{.}
\end{equation}
In the radiative dynamics of the meta-atoms, described by
$\mathcal{C}$, we separate the contributions of
$\mathcal{C}_{\mathrm{SSR}}$ and $i\delta\omega\mathcal{A}$, so that
the matrix $\mathcal{C}_{\mathrm{SSR}}$ describes the collective
dynamics of the metamaterial in the absence of asymmetry (i.e. in an
array where all ASRs are replaced by SSRs), and
$\delta\omega\mathcal{A}$ accounts for the
resonance shifts of the individual meta-atoms due to the asymmetry of
the ASRs.
The diagonal elements of the interaction matrix $\mathcal{C}_{\mathrm{SSR}\,j,j} =
-\Gamma/2$, where
$\Gamma \equiv \Gamma_{\sjE} + \Gamma_{\sjM}$, represent decay rates
and its off-diagonal elements interactions mediated by the scattered
EM field.
The asymmetry in the ASRs shifts the meta-atom resonance frequencies
by $\pm\delta\omega$.
The sign of the frequency shift for a given meta-atom is contained in
the diagonal matrix $\mathcal{A} =
\operatorname{diag}(-1, 1, \ldots, -1, 1)$; the alternating signs of the
elements indicate that the asymmetry shifts the frequencies
of each side of the ASR in opposite directions.
As a result of the incident wave, each element $j$ also experiences a driving $F_j = F_0
\exp[i(\spvec{k}\cdot \spvec{r} - \Delta{}t)]$, $\Delta \equiv \Omega
- \omega_0$, with uniform amplitude $F_0$.

In the following analysis, it is beneficial to consider the collective
modes that are eigenvectors of $\mathcal{C}_{\mathrm{SSR}}$, i.e.,
eigenmodes of a metamaterial in the absence of meta-molecule
asymmetries.
Of particular interest among the collective modes are the PME and PMM
modes with phase-coherent electric and magnetic dipole excitations,
respectively.
The incident wave, whose electric field is parallel to the ASR
electric dipoles, drives the PME mode.
Since the incident wave's magnetic field is perpendicular to
the ASR magnetic dipoles, the PMM mode is not directly driven.
The asymmetry, however, couples the collective modes to each other
in a way similar to how it couples the symmetric and antisymmetric
oscillations of a single isolated ASR (See Appendix~\ref{sec:asymm-induc-coupl}).
The phases and amplitudes of the electric dipoles in the PME
mode closely match those of the magnetic dipoles
in the PMM mode.
Because of this mode matching, the asymmetry couples the PME and PMM
modes more strongly to each other than to any other mode in the system.
We therefore initially ignore coupling of other collective modes to the PME and PMM modes.
(This is later justified by the full numerical calculation and in
Appendix~\ref{sec:asymm-induc-coupl}.)
The dynamics is therefore approximated
by,
\begin{subequations}
  \begin{align}
    \label{eq:c_E_equation}
    \dot{c}_{\sjE} = &\left(-i\delta_{\sjE} -
      \gamma_{\sjE}/2\right) c_{\sjE} -i\delta\omega
    c_{\sjM} + f_{\sjE} \\
    \dot{c}_{\sjM} = & \left(-i\delta_{\sjM} -
      \gamma_{\sjM}/2 \right) c_{\sjM} -i\delta\omega
    c_{\sjE} \textrm{,} \label{eq:c_M_equation}
  \end{align}
  \label{eq:c_E_c_M_equations}
\end{subequations}
where the subscripts $\sjE$ and $\sjM$ refer to PME and PMM modes, respectively
(excitation amplitudes $c_{\sjE,\sjM}$, resonance frequency shifts $\delta_{\sjE,\sjM}$, decay rates
$\gamma_{\sjE,\sjM}$, and driving $f_{\sjE}$).
Equations~\eqref{eq:c_E_equation} and \eqref{eq:c_M_equation} are
similar to those that describe the dynamics of atomic coherences in
EIT \cite{FleischhauerEtAlRMP2005}.
Namely, when the system is driven on resonance with the PMM mode and
$(\delta\omega)^2 \gg  \gamma_\sjM \gamma_\sjE$, the PMM mode is
excited and the asymmetry induced coupling between the PMM and PME
destructively interferes with the driving of the PME mode to prevent
its excitation.

In the calculation of the transmittance and reflectance we consider
the field scattered from the resonators in the
forward, $\unitvec{e}_z$, and backward, $-\unitvec{e}_z$, directions
in the far field. We assume an absorbing planar barrier
is placed around the metamaterial array so that the incident field can
propagate through the array, but not around it, yielding the
diffracted far field  component of the incident field in the forward
direction, $E_\sjI \equiv \unitvec{d} \cdot
\spvec{E}_{\mathrm{I}}(\unitvec{e}_z)$ (See Appendix~\ref{sec:light-scatt-transm}).
Both the incident and the scattered fields
$\spvec{E}_\sjS(\pm \unitvec{e}_z)$ are polarized along the meta-atom electric dipoles.
Therefore, we define the transmittance and reflectance amplitudes as
$T = (E_{\mathrm{I}} +
 \unitvec{d} \cdot \spvec{E}_{\mathrm{S}}(\unitvec{e}_z))/E_{\mathrm{I}}$  and $R =
 \unitvec{d}\cdot\spvec{E}_{\mathrm{S}}(-\unitvec{e}_z)/E_{\mathrm{I}}$
 (See Appendix~\ref{sec:light-scatt-transm}).

We first estimate $R$ and $T$ in a phenomenological model by solving the steady-state response of Eqs~\eqref{eq:c_E_c_M_equations} and assuming a uniformly excited array. This simplified approach
is then compared with a full numerical solution of Eq.~\eqref{eq:ColDyn} that incorporates all
collective modes and the finite-size effects.
In the phenomenological uniform mode approximation (See Appendix~\ref{sec:two-mode-model}), we find
\begin{equation}
  \label{eq:R_amp}
  R =
  \frac{R_0\gamma_{\sjE}/2\left[\gamma_{\sjM}/2-i\left(\Delta - \delta_{\sjM}\right)\right]}
  {(\delta\omega)^2 - \left(\Delta -
      \delta_{\sjE} + i\gamma_{\sjE}/2\right)\left(\Delta
      - \delta_{\sjM} +i\gamma_{\sjM}/2\right)}
  \textrm{ ,}
\end{equation}
and $T = 1 + R$, where $R_0 = -3(\Gamma_{\sjE}/\gamma_{\sjE})/[2\pi(a/\lambda)^2]$
is the reflectance of the system on resonance with the PME mode when
$\delta\omega = 0$, and $\lambda \equiv c/(2\pi\omega_0)$. 
The phenomenological model \eqref{eq:R_amp}
depends on the parameters of the collective modes PME and PMM, $\gamma_{\sjE,\sjM}$ and $\delta_{\sjE,\sjM}$,
that may be calculated numerically (See
Appendix~\ref{subsec:mode-param-estimation}). 
Some example values are given in Table~\ref{tab:PMEPMMProps}.
To illustrate the cooperative nature of CAIT, we here examine the transmission properties of three
different sized arrays: a small (11$\times$11), medium
(41$\times$41), and large (205$\times$205).
All have lattice spacing $a=0.4\lambda$, $\Gamma_{\sjE} = \Gamma_{\sjM}$, and are composed of
ASRs whose meta-atoms are separated by $u = 0.18\lambda$. 

\begin{table}
  \centering
  \begin{tabular}{|rl|c|cc|cc|}
    \hline
    \multicolumn{2}{|c|}{Array Size} & $\,\delta\omega/\Gamma\,$ & $\,\delta_{\sjE}/\Gamma\,$ &
    $\,\gamma_{\sjE}/\Gamma\,$ & $\,\delta_{\sjM}/\Gamma\,$ &
    $\,\gamma_{\sjM}/\Gamma\,$ \\
    \hline \hline
    small:& $11\times{}11$ & $0.1$  & $0.76$ & $1.5$ & $0.57$ & $0.034$ \\
    medium:& $41\times{}41$ & $0.1$  & $0.79$ & $1.5$ & $0.56$ & $3.0 \times
    10^{-3}$ \\
    large:& $205\times{205}$ & $0.02$ & $0.79$ & $1.5$ & $0.56$ & $1.2 \times
    10^{-4}$\\
    \hline
  \end{tabular}
  \caption{The ASR asymmetries and the PME and PMM mode properties of
    the 2D ASR arrays used to demonstrate CAIT.
    The linewidth $\gamma_\sjM$ varies inversely with the size of the
    array.
 }
  \label{tab:PMEPMMProps}
\end{table}

\begin{figure}
  \centering
  \includegraphics[width=0.49\columnwidth]{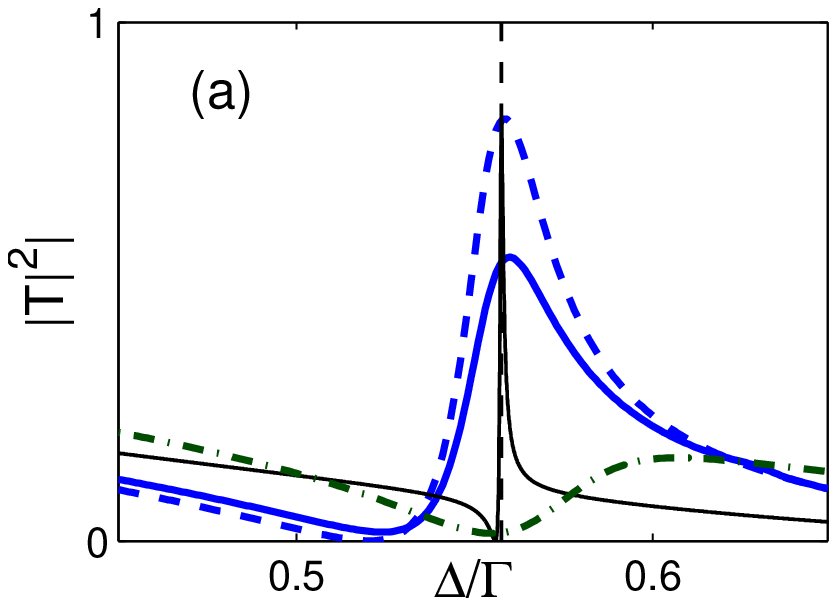}
  \includegraphics[width=0.49\columnwidth]{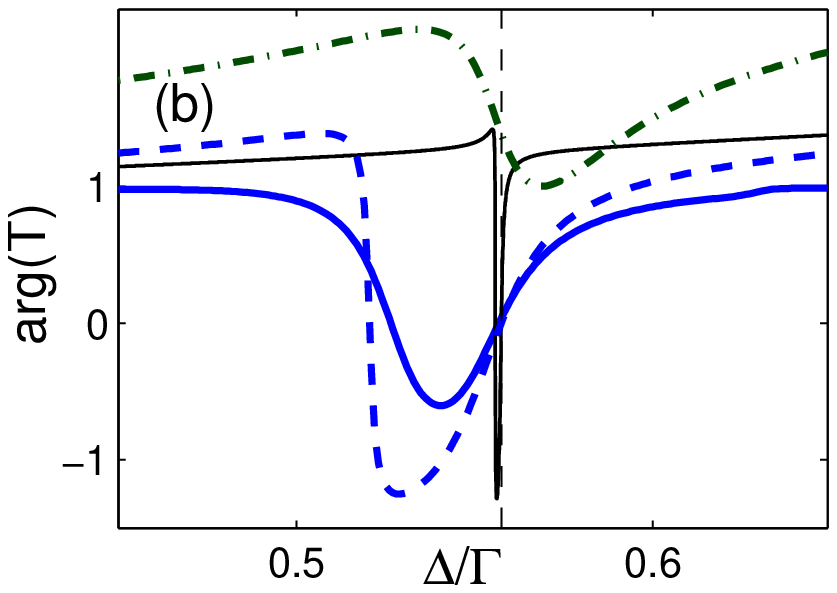}
  \caption{(Color online)
    (a) The intensity transmittance $|T|^2$ and (b) the phase delay
    $\arg (T)$ of the small ASR array in the full model (dot-dashed
    green line), the medium array in the full model (solid blue line),
    the medium array in the uniform mode approximation (dashed blue
    line), and the large array in the uniform mode approximation (thin
    black line).
    Simultaneously increasing the array size and reducing the
    asymmetry narrows the CAIT resonance.
 }
  \label{fig:CAIT_Ideal}
\end{figure}

Figure~\ref{fig:CAIT_Ideal} shows that, in the medium array,
the uniform mode approximation reproduces the qualitative behavior of the full model
[Eq.~\eqref{eq:ColDyn}].
This correspondence indicates that the PME and PMM modes play the
dominant role in governing the array's transmission properties.
The discrepancy arises due to finite-size effects in the full model,
which, for example, allow the excitation of modes other than the PME
and PMM modes.

Comparing the transmission spectra in Fig.~\ref{fig:CAIT_Ideal}, one
finds that the medium and large arrays support CAIT, while the small
array does not.
When $\gamma_{\sjM} \ll (\delta\omega)^2/\gamma_{\sjE}$, as in the medium and large
arrays, excitation of the PME mode is suppressed in a range of frequencies
around the PMM resonance (See Appendix~\ref{sec:two-mode-model}).
This suppression reduces reflection, opening a transparency window.
Equation~\eqref{eq:R_amp} indicates that resonant driving of the PMM
maximizes the intensity transmittance when $|\delta_{\sjE} -
\delta_{\sjM}| \ll \gamma_{\sjE}$.
The quality, or inverse spectral width, of the
resonance increases in proportion to
$(\delta\omega)^{-2}$ (See Appendix~\ref{sec:decr-linew-incr}).
But the condition $\gamma_{\sjM} \ll (\delta\omega)^2/\gamma_{\sjE}$ imposes an
upper bound on the achievable quality, as illustrated by the lack of a
transparency window for the small array.

In contrast to EIT, the decay rate of the cooperative dark mode asymptotically scales as $\gamma_{\sjM}\sim 1/N$
with the number $N$ of ASRs~\cite{JenkinsLineWidthNJP}. This permits one to narrow the transparency window by designing an array with a greater $N$ and reduced $\delta\omega$, even when the constituent resonators
individually would exhibit broad linewidths.
On PMM resonance, Eq.~\eqref{eq:R_amp} implies that the minimum asymmetry required to suppress $R$
below a given level $\delta\omega_{\mathrm{min}}\propto
\sqrt{\gamma_{\sjM}}$ (See Appendix~\ref{sec:decr-linew-incr}).
Hence the maximum attainable quality factor [$\propto (\delta\omega)^{-2}$] of the
transparency window increases in proportion to $N$ and is eventually only limited by nonradiative losses, resulting in very sharp resonances with high modulation depths.
For example, from the asymptotic expressions of $\gamma_{\sjE,\sjM}$ and
$\delta_{\sjE,\sjM}$ (See Appendix~\ref{subsec:mode-param-estimation}) we can deduce that simultaneously quintupling the
side lengths of an array and reducing $\delta\omega$ by a factor of
five narrows the resonance from about $\Gamma/40$
to $\Gamma/1000$ (Appendix~\ref{sec:transResChar}), while maintaining the peak
transmittance (Fig.~\ref{fig:CAIT_Ideal}).

The sharp transmission resonance exhibits a considerable phase delay $\varphi(\Delta) \equiv \arg
(T(\Delta))$ on $\Delta$. According to numerics a pulse resonant on the PMM mode passing through the
41$\times$41 sample would experience a group delay of $\tau_g \equiv
d\varphi/d\Delta|_{\Delta = \delta_{\rmM}} \approx 47/\Gamma$. The
delay is further enhanced in the large array owing to linewidth
narrowing and we estimate $\tau_g \approx 1600/\Gamma$ in
the phenomenological model.

\begin{figure}
 \centering
 \includegraphics[width=0.49\columnwidth]{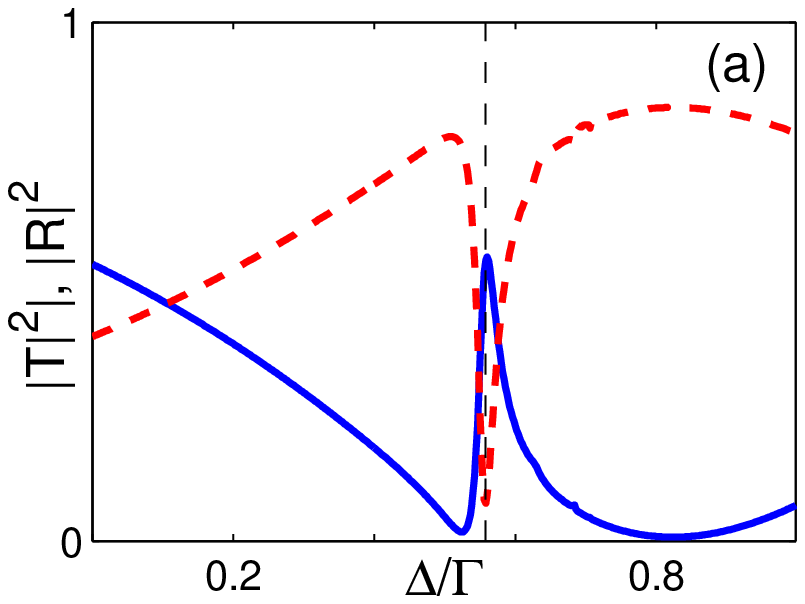}
 \includegraphics[width=0.49\columnwidth]{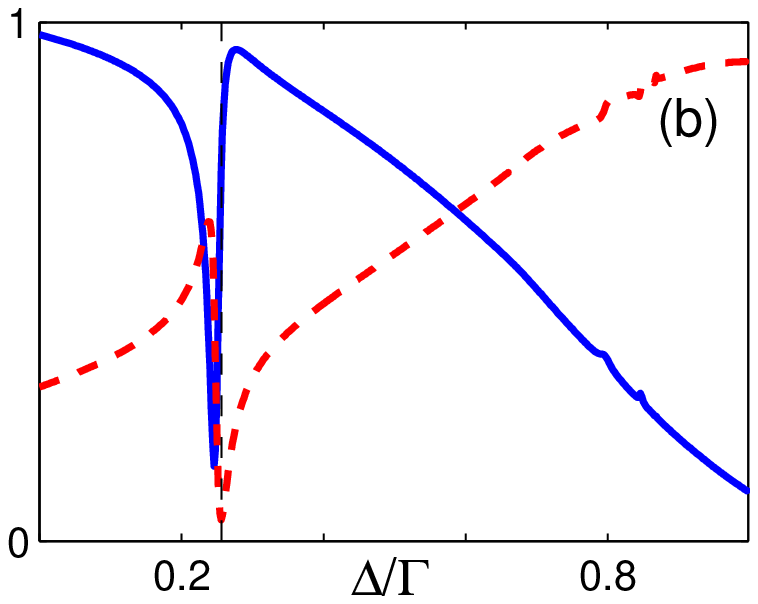}
 \caption{(Color online) The effect of reconfiguring the metamaterial geometry on the
   transmission resonance. The intensity transmittance (solid line) and
   intensity reflectance (dashed line) are calculated for a (a) $41\times 41$
   2D array and (b) a lattice in which the two sublattices are shifted
   by $\delta\spvec{R} =0.1\unitvec{e}_z$.
   Calculations account for excitation of and scattering from all
   collective modes.
   The vertical dashed lines indicate the PMM resonance.
}
 \label{fig:Shifted}
\end{figure}

Dynamically reconfiguring the metamaterial geometry \cite{PryceEtAlNanLett2010,OuEtAlNatNan2013} provides an active
control mechanism for the transparency.
To illustrate this, we split the medium array into two interleaved
sublattices with lattice vectors $\spvec{a}_1 = 2a \unitvec{e}_x$ and
$\spvec{a}_2 = a\unitvec{e}_y$.
The lattices are displaced from one and other by $a\unitvec{e}_x +
\delta\spvec{R}$ so that for $\delta\spvec{R}
=0$, the ASRs form a square lattice.
Figure~\ref{fig:Shifted} shows how distorting the
lattice alters the transmission resonance.
Displacing the sublattices by $\delta\spvec{R} =
-0.1\lambda \unitvec{e}_z$ generates a relative shift of about $0.5\Gamma$ between the  PME and PMM resonances,
almost entirely eliminating the transparency.
A fast control of metamaterial arrays~\cite{OuEtAlNatNan2013}, together with the sensitivity of cooperative resonances to
the specific resonator configuration, could potentially open possibilities for
stopped pulse and light storage applications~\cite{LiuEtAlNature2001}.

In conclusion, we proposed a controllable mechanism to produce
a cooperative transmission resonance CAIT.
Whereas standard EIT can be described by
independent emitters, CAIT relies on a cooperative response
of the metamaterial.
A transmission resonance forms when a subradiant collective mode,
acting as a dark state, is excited at the expense of the mode that
most efficiently couples to an incident EM field.
Since the lifetime of the dark PMM mode increases with size of the
array \cite{JenkinsLineWidthNJP}, the attainable quality of the resonance scales in proportion
with the number of resonators in the metamaterial.
For large arrays, only nonradiative decay, which could be
incorporated into the analysis using a phenomenological parameter
\cite{JenkinsLineWidthNJP}, limits the attainable quality factor of
the resonance. 
In low-loss materials, such as superconducting metamaterial arrays~\cite{RicciEtAlAPL2005,Lazarides,Du,GuEtAlAPL2010,Rakhmanov,SavinovEtAlSciRep2012}, the nonradiative decay,
however, can be suppressed.

\begin{acknowledgements}
  This work was supported by the EPSRC and the Leverhulme Trust. We
  would like to thank N. Papasimakis, V. Fedotov, V. Savinov, E. Plum, M. D. Lee,
  M. O. Borgh and N. I. Zheludev for discussions.
\end{acknowledgements}

\appendix

\section{Asymmetric split rings in the metamaterial array}
\label{sec:descr-asymm-split}

We base our analysis of the CAIT response on a general formalism we
developed \cite{JenkinsLongPRB} to describe collectively interacting
metamaterial arrays of magnetodielectric resonators.
When applied to 2D arrays of strongly coupled asymmetric split rings
(ASRs), this model yielded an excellent agreement
\cite{JenkinsLineWidthNJP} with the experimental transmission
resonance measurements in Ref.~\cite{JenkinsLongPRB}.

To show how CAIT arises from cooperative phenomena, we consider an
ensemble of $N$ identical ASRs, each composed of two
circular arcs, or meta-atoms (see Fig.~\ref{fig:LevelStructure}), lying on
or near (within a tenth of a wavelegth) of the $z=0$ plane.
The meta-atoms are labeled by indices $j$ ($j = 1,\ldots,2N$) such
that ASR $\ell$ ($\ell = 1,\ldots,{N}$) comprises meta-atoms $2\ell -1$
and $2\ell$.
Although, in general, each meta-atom $j$ occupies an area comparable to that
of the metamaterial unit-cell \cite{FedotovEtAlPRL2010}, we
can produce the qualitative behavior of an ASR by approximating
meta-atom $j$ as a point source for the electromagnetic (EM) field at position
$\spvec{r}_j$, where the arcs of an ASR are separated by
$\spvec{u} \equiv \spvec{r}_{2\ell} - \spvec{r}_{2\ell-1} \equiv u\unitvec{u}$
\cite{JenkinsLineWidthNJP}.
In this work, we assume the ASRs are oriented such that $\unitvec{u} = \unitvec{e}_x$.

We assume each meta-atom $j$ supports a single mode of current oscillation that
behaves as an effective $LC$ circuit with resonance frequency $\omega_j$~\cite{JenkinsLongPRB}.
If the split rings were symmetric, the individual meta-atoms would have
identical resonance frequencies $\omega_j = \omega_0$.
An asymmetry in the arc lengths shifts the meta-atom
resonance frequencies by $\delta\omega$ so that for ASR $\ell$
\begin{subequations}
  \label{eq:meta_at_freqs}
  \begin{eqnarray}
    \omega_{2\ell-1} = \omega_0 - \delta\omega \textrm{ ,}\\
    \omega_{2\ell} = \omega_0 + \delta\omega \textrm{ .}
  \end{eqnarray}
\end{subequations}
Oscillating currents in meta-atom $j$ produce an electric dipole
$\spvec{d}_j(t) = d_j(t)\unitvec{d}$  and
magnetic dipole $\spvec{m}_j(t) = m_j(t)\unitvec{m}_j$, where $d_j$
and $m_j$ are the electric and magnetic dipole amplitudes, respectively.
The resulting electric and magnetic dipole radiation damps current
oscillations in the meta-atom at rates $\Gamma_{\sjE}$ and $\Gamma_{\sjM}$, respectively \cite{JenkinsLongPRB}.
All ASRs in the array have the same orientation so that
every meta-atom's electric dipole points in the direction
$\unitvec{d} \equiv \unitvec{e}_y$.
The magnetic dipole directions $\unitvec{m}_j$, on the other hand,
are oriented so that identical current flows in an ASR produce opposite magnetic
dipoles in its constituent meta-atoms.
That is, the right ($j=2\ell$) and left ($j=2\ell-1$) meta-atoms in
ASR $\ell$ have magnetic dipole orientations
$\unitvec{m}_{2\ell} = -\unitvec{m}_{2\ell-1} \equiv \unitvec{m} =
\unitvec{u} \times \unitvec{d}=\unitvec{e}_z$.
In this way, identical current flows in an ASR's meta-atoms produce an
electric dipole, while equal and opposite flows produce a magnetic
dipole.
The dynamics of the oscillating dipoles within each meta-atom $j$ are
described by the slowly varying normal variable \cite{JenkinsLongPRB}
\begin{equation}
  \label{eq:b_def}
  b_j(t) = \sqrt{\frac{k_0^3}{12\pi\epsilon_0}}\left(
    \frac{d_j(t)}{\sqrt{\Gamma_{\sjE}}} + i
    \frac{m_j(t)}{c\sqrt{\Gamma_{\sjM}}}\right) \textrm{,}
\end{equation}
where $k_0 \equiv \omega_0/c$.

\section{Asymmetry induced coupling of collective modes}
\label{sec:asymm-induc-coupl}

In this appendix, we show how the asymmetry in the ASRs couples collective modes of the system.
Of particular importance to CAIT are the phase matched electric (PME)
mode, in which all electric dipoles oscillate in phase, and the phase
matched magnetic (PMM) mode, in which all magnetic dipoles oscillate in phase.
The PME mode is phase-matched with EM plane-waves propagating perpendicular to
the array and can be easily excited by an incident field. For ASRs in a square lattice, 
we show that the PME and PMM modes couple almost exclusively to each other.
This exclusivity yields the simplified response of the PME and PMM
modes described by Eqs.~\eqref{eq:c_E_c_M_equations}.

To describe the dynamics of the metamaterial comprising an ensemble of
ASRs, we employ the column vector of normal variables,
\begin{equation}
  \colvec{b}(t) = \left(
    \begin{array}{c}
      b_1 (t) \\
      b_2 (t) \\
      \vdots \\
      b_{2N-1}(t) \\
      b_{2N}(t)
    \end{array}
  \right) \textrm{ .}
\end{equation}
We associate collective modes of the system with eigenvectors of
$\mathcal{C}_{\mathrm{SSR}}$, i.e., eigenmodes of a metamaterial
where all ASRs are replaced by symmetric split rings (SSRs).
The $n^{\mathrm{th}}$ collective mode corresponds to the
eigenvector $\colvec{v}_n$ of the interaction matrix
$\mathcal{C}_\mathrm{SSR}$ 
normalized such that $\colvec{v}_m^T\colvec{v}_n =
  \delta_{mn}$.
The state
$\colvec{b}(t)$ can be expanded as  
\begin{equation}
  \label{eq:colModes}
  \colvec{b}(t) = \sum_n c_n(t) \colvec{v}_n \textrm{,}
\end{equation}
where $c_n(t) \equiv \colvec{v}_n^T \colvec{b}(t)$ is the amplitude of the
collective mode $n$.
By expressing $\colvec{b}$ in the basis of collective modes $\{\colvec{v}_n : n=1,\ldots,2N\}$, we find that the asymmetry results in the evolution of the amplitudes,
\begin{equation}
  \label{eq:ModeAmp_eq_m}
  \dot{c}_n = \left(-i\delta_n - \gamma_n/2\right)c_n - i\delta\omega \sum_{m} c_m
  \left(\colvec{v}_n^T \mathcal{A} \colvec{v}_m \right) + f_n \, \text{,}
\end{equation}
where $\delta_n$ and $\gamma_n$ are the frequency shift and decay rate
of mode $n$, corresponding to the $n^{\mathrm{th}}$ eigenvalue of $\mathcal{C}_{\mathrm{SSR}}$.
The matrix element $\left(\colvec{v}_n^T \mathcal{A} \colvec{v}_m \right)$ represents the coupling between modes $m$ and $n$, caused by the asymmetry.

We identify the PME (PMM) mode as the eigenvector of
$\mathcal{C}_\mathrm{SSR}$ that most
resemble all electric (magnetic) dipoles oscillating
in phase with equal amplitudes~\cite{JenkinsLineWidthNJP}.
The coupling between the PME amplitude $c_\sjE$ and the PMM
amplitude $c_\sjM$, given by $\colvec{v}_{\sjM}^T \mathcal{A}
\colvec{v}_{\sjE}$, can dominate over their coupling to other modes
for particular arrangements of ASRs.
For example, in the $41\times 41$ (medium) array considered in the main text, the asymmetric
coupling coefficient is $\left|\colvec{v}_{\sjM}^T \mathcal{A}
\colvec{v}_{\sjE}\right| \approx 0.9993$ and the maximal coupling of the PME and
PMM modes to other collective modes is $\colvec{v}_{n}^T \mathcal{A}
\colvec{v}_{\sjE}, \colvec{v}_{n}^T \mathcal{A}
\colvec{v}_{\sjM} < 0.04$ for $n\notin \{\sjE,\sjM\}$.
The PME and PMM modes thus obey the simplified dynamics given by
Eqs.~\eqref{eq:c_E_c_M_equations}.

In sum, if the incident field does not directly drive any other modes,
the near exclusive coupling of the PME and PMM modes to each other
forms an effective two-mode system for the metamaterial. This further justifies the
use of the phenomenological model in the main text.

\section{Scattered light, transmittance and reflectance}
\label{sec:light-scatt-transm}

To analyze the light scattering properties of the array, including its
reflectance and transmittance, we consider a  plane wave of frequency
$\Omega$ impinging on the array with positive frequency component 
\begin{equation}
  \spvec{E}^+_{\mathrm{in}}(\rv,t) = \unitvec{d}\mathcal{E}e^{i (kz
    -\Omega t)}\,, 
\end{equation}
where $k\equiv\Omega/c$, and $\mathcal{E}$ is the electric field
amplitude.
The incident wave is detuned from the central meta-atom resonance
frequency by $\Delta \equiv \Omega-\omega_0$.
According to Eq.~\eqref{eq:ColDyn}, driving by the incident wave induces a
cooperative response of the array.
The
oscillating electric and magnetic dipoles emit a scattered field
$\spvec{E}_{\sjS}$.

So as to consider the transmittance through the metamaterial array, we assume a
2D barrier is placed around the array in the $z=0$ plane
so that fields can propagate through the array, but not around it.
We denote the incident field that would diffract through the barrier
if the array were not present as
$\spvec{E}_\sjI$.
The field transmitted through the array ($z>0$) is then $\spvec{E}_\sjT \equiv
\spvec{E}_\sjI + \spvec{E}_{\sjS}$.
On the other hand, the field reflected from the array ($z<0$) is determined
only by the scattered field $\spvec{E}_{\sjS}$.

For simplicity, we analyze the diffracted and scattered fields in the
far field, observed a distance $R$ from the
metamaterial much greater than the spatial extent of the array.
Here, we
define the far field amplitude $\spvec{E}(\unitvec{k},\Omega) $ of the electric field along the
direction $\unitvec{k}$ (with wavevector $\spvec{k}=k\unitvec{k}$)
such that the positive component of the electric field at $R\unitvec{k}$,
\begin{equation}
  \spvec{E}^+(R\unitvec{k},t) \approx \frac{e^{i(kr-\Omega t)}}{kR}
  \spvec{E}(\unitvec{k},\Omega) \textrm{.}
\label{eq:farFieldDef}
\end{equation}
According to Fraunhofer diffraction \cite{Zangwill}, the diffracted incident field then reads
\begin{equation}
  \spvec{E}_{\sjI}(\unitvec{k}) =-ik^2\left(\unitvec{k} \times
    \unitvec{e}_x\right)
  \frac{1}{2\pi} \int_{A} dxdy\, \mathcal{E} e^{-i\spvec{k}\cdot
    \spvec{r}_\perp}  \text{,} \label{eq:E_I}
\end{equation}
where $A$ is the area of the aperture, and $\spvec{r}_\perp \equiv
x\unitvec{e}_x + y\unitvec{e}_y$.
The barrier that surrounds an $N_x \times N_y$
lattice centered at the origin contains an $N_x a \times N_y a$ rectangular
aperture where $a$ is the lattice spacing.
From Eq.~\eqref{eq:E_I}, the forward diffracted component of the
incident field through such a barrier is therefore given by
\begin{equation}
  \spvec{E}_{\sjI}(\unitvec{k}) = -i
  \frac{N\mathcal{E}\left(ka\right)^2}{2\pi} \left(\unitvec{k}\times
    \unitvec{e}_x\right) \prod_{j=x,y} \sinc\left(\frac{N_jk_ja}{2}\right) \text{,} \label{eq:E_I2}
\end{equation}

The scattered fields, on the other hand, result from electric and
magnetic dipole radiation emitted by the meta-atoms.
Their far-field components are given by \cite{Zangwill}
\begin{equation}
  \label{eq:E_S_far_Field}
  \spvec{E}_{\sjS}(\unitvec{k},\Omega) = \frac{k^3}{4\pi\epsilon_0}
  \unitvec{k} \times \sum_{j=1}^{2N}
  \left[\left(\tilde{\spvec{d}}_j \times \unitvec{k}\right)
    -
    \frac{\tilde{\spvec{m}}_j}{c} \right]
  e^{-i\spvec{k}\cdot \spvec{r}_j} \textrm{,}
\end{equation}
where $\tilde{\spvec{d}}_j \equiv e^{i\Omega t}\spvec{d}_j^+$ and
$\tilde{\spvec{m}}_j \equiv e^{i\Omega t}\spvec{m}_j^{+}$ are the
slowly varying
electric and magnetic dipoles, respectively, of meta-atom $j$.
From Eq.~\eqref{eq:b_def}, which relates the electric and magnetic
dipoles to the amplitudes $b_j$, we can express the scattered
far field amplitude as
\begin{align}
  \label{eq:E_S_far_field_in_bs}
  \spvec{E}_\sjS(\unitvec{k},\Omega) =
  - & \sqrt{\frac{3k^3}{16\pi\epsilon_0}}  \unitvec{k} \times 
   \sum_{j=1}^{2N} \left[\sqrt{\Gamma_{\sjE}} \left(\unitvec{k}
      \times \unitvec{d}\right) \right.  \nonumber \\
  & \left. +i (-1)^j
    \sqrt{\Gamma_{\sjM}}\unitvec{m}\right]  b_j(\Delta)
  e^{-i\spvec{k}\cdot \spvec{r}_j}\textrm{.}
\end{align}

In this work, we define the reflectance and transmittance in terms of the
backward ($\unitvec{k}=-\unitvec{e}_z$) and forward scattered
($\unitvec{k}=\unitvec{e}_z$) fields, respectively.
Since the electric dipoles and incident field are oriented along
$\unitvec{d}=\unitvec{e}_y$, and the magnetic dipoles are parallel to $\unitvec{e}_z$,
the forward and backward scattered fields will
be polarized along $\unitvec{d}$. We therefore define the forward
transmittance and backward reflectance as
\begin{subequations}
  \begin{eqnarray}
    R &\equiv& \frac{\unitvec{d}\cdot
      \spvec{E}_\sjS(-\unitvec{e}_z,\Omega)}{ \unitvec{d} \cdot \spvec{E}_\sjI(\unitvec{e}_z)} \textrm{,}  \label{eq:R_def}\\
    T & \equiv
    &\frac{\unitvec{d}\cdot\left(\spvec{E}_\sjI(\unitvec{e}_z) +
        \spvec{E}_\sjS(\unitvec{e}_z,\Omega)\right)}{ \unitvec{d} \cdot
      \spvec{E}_\sjI(\unitvec{e}_z)} \textrm{,} \label{T_def}
  \end{eqnarray}
\end{subequations}
One could similarly define the reflectance and transmittance
amplitudes in terms
of the incident and scattered fields integrated over some solid angle
about $\pm\unitvec{e}_z$.
We have checked that in the numerical simulations discussed in the
text, integration over a sufficiently small solid angle does not alter
the phase or amplitude dependence of the transmitted and reflected
fields.

\section{Characterizing a transmission resonance}
\label{sec:transResChar}

For a given transmittance amplitude $T(\Delta)$, a transmission
resonance occurring at $\delta_{\sjT}$ is characterized by peak
transmittance $|T(\delta_{\sjT})|^2$, resonance width $w$,
and group delay $\tau_g$ of a resonant pulse passing through the
metamaterial.

To determine the resonance width, we can approximate the intensity
transmittance $|T(\Delta)|^2$ to second order in
$\Delta-\delta_{\sjT}$ by a Gaussian of height
$|T(\delta_{\sjT})|^2$, and full width at half max (FWHM)
$w$.
Comparing the Taylor expansions of the two, we have
\begin{align}
  \label{eq:3}
  |T(\Delta)|^2 &\approx |T(\delta_{\sjT})|^2+
  \frac{1}{2}\left.\frac{d|T|^2}{d\Delta^2}\right|_{\Delta=\delta_{\sjT}}
  (\Delta-\delta_{\sjT})^2 \nonumber\\
  & \approx |T(\delta_{\sjT})|^2 \exp\left(-\log
    2\frac{4\left(\Delta-\delta_{\sjT}\right)^2}{w^2}\right) \textrm{,}
\end{align}
where the FWHM of the Gaussian is
\begin{equation}
  \label{eq:w_def}
  w = \sqrt{-\frac{8(\log 2)
    |T(\delta_{\sjT})|^2}{\left. d|T|^2/d\Delta^2\right|_{\Delta=\delta_{\sjT}}}
  } \textrm{.}
\end{equation}
In this work, we estimate the width of the resonance to be the FWHM
$w$ of the Gaussian that best fits the intensity transmittance near
the peak.

The group delay $\tau_g$ of a resonant pulse passing through the
metamaterial is determined by the phase of the transmittance amplitude
$\arg(T)$. Specifically,
\begin{equation}
  \label{eq:tau_g}
  \tau_g \equiv \left.\frac{d}{d\Delta} \arg
    T(\Delta)\right|_{\Delta=\delta_{\sjT}}  =
  -i\frac{1}{T}\left.\frac{dT}{d\Delta}\right|_{\Delta =\delta_{\sjT}}
\end{equation}

\section{The uniform mode approximation for scattering from a planar array}
\label{sec:two-mode-model}

We use the phenomenological two-mode model [Eq.~\eqref{eq:c_E_c_M_equations}]
to estimate the reflectance and transmittance from a uniformly
excited planar array of ASRs in which case we neglect boundary effects.
We refer to this assumption as the uniform mode approximation.
For the uniform PME and PMM modes the frequency shifts
$\delta_{\sjE,\sjM}$, collective decay rates
$\gamma_{\sjE,\sjM}$, and asymmetry $\delta\omega$
fully describe the cooperative response to an incident plane wave.

We consider an $N_x \times N_y$ ($N=N_xN_y$) square lattice of ASRs with lattice spacing $a$ and lattice vectors $\spvec{a}_1 = a \unitvec{e}_x$ and $\spvec{a}_2 = a\unitvec{e}_y$.
In the uniform mode approximation, The PME and PMM modes consist of
uniformly excited electric and magnetic dipoles, respectively.
Explicitly, these modes correspond to the vectors
\begin{equation}
  \label{eq:UMA_PME_PMM}
  \colvec{v}_{\sjE} =   \frac{1}{\sqrt{2N}}\left(
    \begin{array}{c}
      1 \\
      1 \\
      \vdots \\
      1 \\
      1
    \end{array}
  \right), \quad
  \colvec{v}_{\sjM} =
  \frac{1}{\sqrt{2N}}
  \left(
    \begin{array}{r}
      -1 \\
      1 \\
      \vdots \\
      -1 \\
      1
    \end{array}
  \right) \textrm{.}
\end{equation}
The alternating signs in $\colvec{v}_{\sjM}$ indicate that the currents in each
ASR flow out of phase with each other in the PMM mode, while in the
PME mode all currents flow in phase.
In the uniform mode limit, an incident plane wave propagating perpendicular to the array can only
drive the PME mode, while the asymmetry couples the PME and PMM
modes only to each other.
As such, the metamaterial response to a field of frequency
$\Omega=\omega_0 + \Delta$ is given by the Fourier components of the
mode amplitudes as
\begin{equation}
  \label{eq:2}
  \colvec{b}(\Delta) = c_{\sjE}(\Delta) \colvec{v}_{\sjE} + c_{\sjM}(\Delta)
  \colvec{v}_{\sjM} \text{.}
\end{equation}

\subsection{The scattered field in the far field from the steady state
metamaterial response}
\label{subsec:scattered-field}

The scattered EM fields are generated by the excitations of the resonators. Each resonator acts as a source
of scattered radiation and the regular metamaterial array produces a field pattern of a diffraction grating.  
In the uniform approximation the excitations are described by Eq.~\eqref{eq:2}. The sum over the meta-atoms
in the scattered field expression~\eqref{eq:E_S_far_field_in_bs} therefore considerably simplifies. We obtain in the limit $k \approx k_0$
\begin{eqnarray}
  \label{eq:E_S_in_c_E}
  \lefteqn{E_\sjS(\unitvec{k},\Omega) = - \sqrt{\frac{3Nk^3}{8\pi\epsilon_0
      }}  \mathcal{D}(\spvec{k})\unitvec{k}\times}
  \nonumber \\
  & &
  \quad\left[\left(\unitvec{k} \times \unitvec{d}\right)
    \sqrt{\Gamma_\sjE} g_\sjE(\unitvec{k}) + i \unitvec{m}\sqrt{\Gamma_\sjM}
    g_\sjM(\unitvec{k})\right] \textrm{.}
\end{eqnarray}
The scattered field is modulated by the diffraction pattern of $N$ unit-cell resonators
$\mathcal{D}(\unitvec{k})= \sum_\ell e^{-i\spvec{k}\cdot \spvec{r}_\ell}/N$, where the summation runs over
all ASRs $\ell$ at positions $\spvec{r}_\ell$. In the studied system we obtain the familiar field amplitude of a 2D square array of $N_x\times N_y$ diffracting apertures
\begin{equation}
  \label{eq:ddelta_k}
  \mathcal{D}(\unitvec{k}) = \frac{\sin(N_xk_xa/2)\sin(N_y k_y
  a/2)}{N\sin(k_xa/2)\sin(k_ya/2)} \textrm{.}
\end{equation}
Owing to the subwavelength lattice spacing $a$, only the zeroth order Bragg peak ($k_x=k_y=0$)
exists. The cone of the emitted radiation in the forward and backward directions $\pm\unitvec{e}_z$ narrows
as a function of the number of unit-cell resonators $N$.

In Eq.~\eqref{eq:E_S_in_c_E}, $g_\sjE(\unitvec{k})$ and $g_{\sjM}(\unitvec{k})$ are, respectively,
proportional to the electric and magnetic dipole
emission of an ASR along direction $\unitvec{k}$. 
They are given
in terms of collective mode amplitudes by
\begin{subequations}
  \label{eq:g_E_and_g_M}
  \begin{align}
    g_{\sjE}(\Delta,\unitvec{k}) &=
    c_{\sjE}\cos\left(\frac{\spvec{k} \cdot
        \spvec{u}}{2}\right) -i c_\sjM
    \sin\left(\frac{\spvec{k} \cdot \spvec{u}}{2}\right) \textrm{,} \\
    g_{\sjM}(\Delta,\unitvec{k}) &=
    c_{\sjM}\cos\left(\frac{\spvec{k} \cdot
        \spvec{u}}{2}\right) -i c_\sjE
    \sin\left(\frac{\spvec{k} \cdot \spvec{u}}{2}\right) \textrm{.}
  \end{align}
\end{subequations}
In the limit $|\spvec{k} \cdot \spvec{u}|\ll 1$, the scattered electric (magnetic) dipole radiation is almost
solely generated by the PME (PMM) mode. The small mixing of these two contributions results from the finite separation $\spvec{u}$ of the two meta-atoms in each unit-cell resonator.

We calculate the scattered fields from the steady-state
solution of Eqs.~\eqref{eq:c_E_c_M_equations},
\begin{subequations}
\begin{align}
  \label{eq:c_E_ss}
  c_\sjE =& -i\frac{Z_{\sjM}(\Delta)}{(\delta\omega)^2 -
    Z_{\sjE}(\Delta)Z_{\sjM}(\Delta)} f_\sjE  \textrm{,} \\
  c_\sjM =& \frac{\delta\omega}{Z_\sjM(\Delta)} c_\sjE \textrm{,}
\end{align}
\end{subequations}
where for each mode $n$, we have defined 
\begin{equation}
  \label{eq:ZDef}
  Z_n(\Delta) \equiv \Delta - \delta_n + i \gamma_{n} \textrm{.}
\end{equation}
The PME and PMM amplitudes are both proportional to the driving
$f_{\sjE}(\Delta)$, which is given by
\begin{equation}
  \label{eq:PME_Driving}
  f_\sjE(\Delta) =  i\sqrt{\frac{6\pi\epsilon_0}{k^3}}\sqrt{\Gamma_{\sjE}N} \mathcal{E} \textrm{ .}
\end{equation}

Having solved the steady-state response of PME mode amplitude 
[Eq.~\eqref{eq:c_E_ss}] and the scattered fields emitted by an
excited PME mode [Eq.~\eqref{eq:E_S_in_c_E}], one finds that
the incident plane wave produces the forward and backward scattered
fields
\begin{equation}
  \spvec{E}_\sjS(\pm \unitvec{e}_z,\Omega) = \unitvec{d}
    \frac{3N\Gamma_{\sjE}}{2} \mathcal{E}
  \frac{Z_{\sjM}(\Delta)}{(\delta\omega)^2 -
    Z_{\sjE}(\Delta)Z_{\sjM}(\Delta)}  \textrm{,}   \label{eq:E_S_E_in}
\end{equation}

\subsection{Reflectance and transmittance}
\label{subsec:refl-trans}

One obtains the transmittance and reflectance by comparing the
scattered fields to the forward propagating component of the incident
field, $\spvec{E}_\sjI(\unitvec{e}_z)$ [Eq.~\eqref{eq:E_I}].
The reflectance associated with the backward scattered field
[Eq.~\eqref{eq:R_def}] and the transmittance of the forward scattered
field [Eq.~\eqref{T_def}], in the uniform mode approximation, are
given by
\begin{align}
  \label{eq:R_amp_app}
  R &=
  \frac{R_0\gamma_{\sjE}/2\left[\gamma_{\sjM}/2-i\left(\Delta - \delta_{\sjM}\right)\right]}
  {(\delta\omega)^2 - \left(\Delta -
      \delta_{\sjE} + i\gamma_{\sjE}/2\right)\left(\Delta
      - \delta_{\sjM} +i\gamma_{\sjM}/2\right)}
   \textrm{ ,}\\
    T & = 1 + R \textrm{ ,} \label{eq:T_amp}
\end{align}
where
\begin{equation}
  R_0 = -\frac{3(\Gamma_{\sjE}/\gamma_{\sjE})}{2\pi(a/\lambda)^2} \label{eq:R0}
\end{equation}
is the reflectance of the system on resonance with the PME mode when
the split rings are symmetric ($\delta\omega=0$), and $\lambda \equiv
c/(2\pi\omega_0)$.

Equation~\eqref{eq:c_E_ss} indicates that, when cooperative effects reduce
$\gamma_{\sjM}$ far below $(\delta\omega)^2/\gamma_{\sjE}$, a field
resonant on the PMM mode does not excite the PME mode.
Rather, the PMM mode is excited, and the asymmetry induced coupling
between the PME and PMM modes destructively interferes with the
driving of the PME mode by the incident field.
The PME mode remains unexcited, and the scattered
field and reflection are suppressed as indicate by
Eq.~\eqref{eq:R_amp_app}.
A transmission resonance therefore forms when $\gamma_\sjM \ll
(\delta\omega)^2/\gamma_\sjE$.

\subsection{Collective mode resonance linewidths and line shifts}
\label{subsec:mode-param-estimation}

\begin{figure}
  \centering
  \includegraphics[width=0.49\columnwidth]{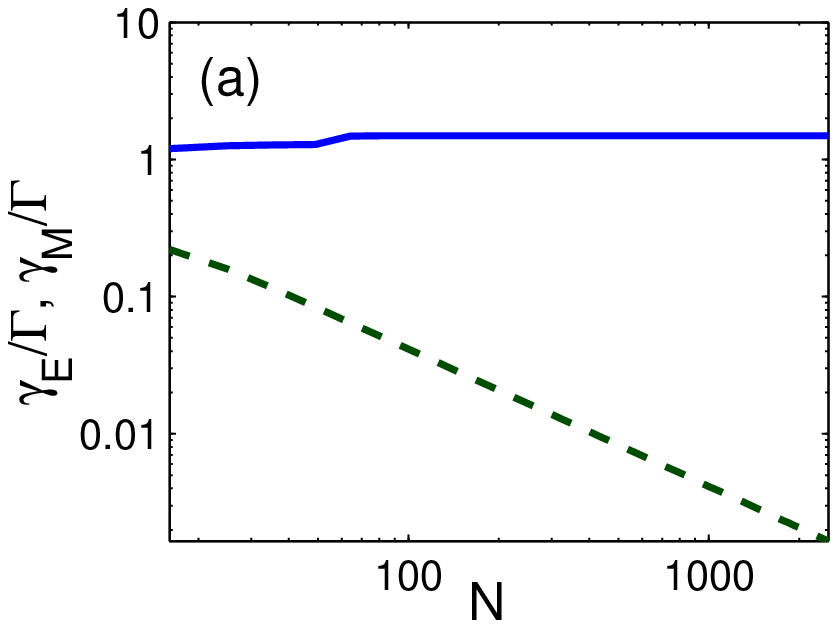}
  \includegraphics[width=0.49\columnwidth]{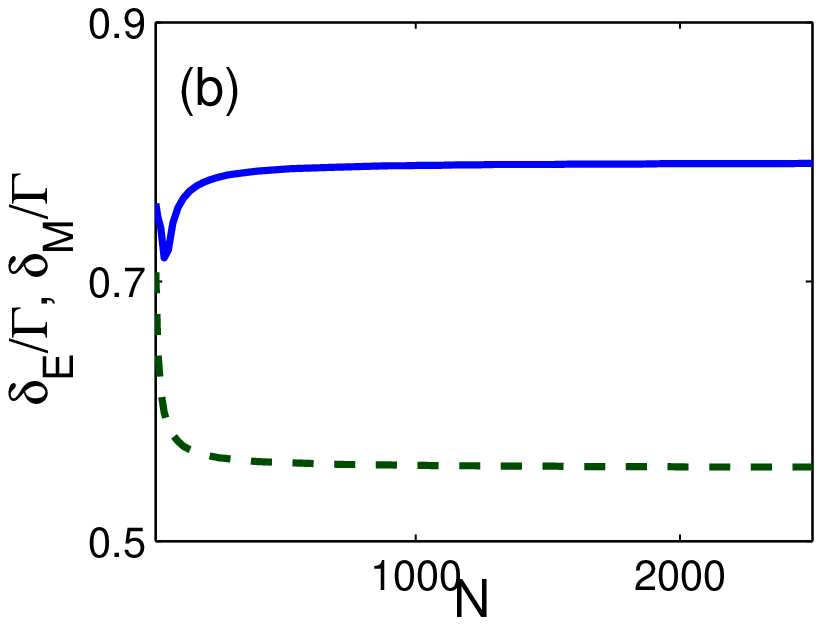}
  \caption{The collective decay rates (a) and collective resonance
    frequency shifts (b) of the PME (solid blue lines) and PMM (dashed green
    lines) modes as functions of $N$ the number of ASRs in a square
    lattice ($N_x=N_y$; $a=0.4\lambda$; $\Gamma_{\sjE}=\Gamma_{\sjM}$;
    $u=0.18\lambda$).  
    The decay rate $\gamma_\sjE$ of the PME mode and the shifts
    $\delta_\sjE$ and $\delta_\sjM$ asymptotically approach constant
    values for sufficiently large $N$.  The decay rate of the PMM mode
    $\gamma_{\sjM}\propto 1/N$ for large $N$.
  }
  \label{fig:gamDelt_v_N}
\end{figure}

To determine the EM response of the array in the uniform mode
approximation, one only needs in Eq.~\eqref{eq:R_amp_app} collective line shifts 
$\delta_{\sjE,\sjM}$ and linewidths $\gamma_{\sjE,\sjM}$ of
PME and PMM modes. In the case of cooperative interactions, these depend on
the number of resonators $N$ in the system. 
In Fig.~\ref{fig:gamDelt_v_N}, we show numerically calculated $\delta_{\sjE,\sjM}$
and $\gamma_{\sjE,\sjM}$ as a function of $N$. 
These are evaluated by
diagonalizing $\mathcal{C}_{\mathrm{SSR}}$ appearing in Eq.~\eqref{eq:ColDyn}. 
Here $\delta_{\sjE,\sjM}$ and $\gamma_{\sjE}$ rapidly reach their approximate 
asymptotic values for array sizes around $N\simeq1000$. 
Identifying the asymptotic behavior allows an efficient calculation of the collective mode 
parameters, the transmittance, and reflectance in the phenomenological uniform
model even for large arrays.

The asymptotic value of $\gamma_{\sjE}$ can also be determined analytically in an infinite
array of SSRs ($\delta\omega=0$) with
subwavelength lattice spacing $a$.
In such a system, the PME mode emits only in the forward and backward
directions, corresponding to the zeroth order diffraction peak.
In the absence of Ohmic losses, energy conservation
therefore requires $|R|^2 + |T|^2=1$ in an infinite array.
Therefore, according to Eqs.~\eqref{eq:R_amp_app} and \eqref{eq:T_amp}, an incident wave
resonant on the PME mode would experience a reflectance amplitude $R_0
= -1$.
At the same time Eq.~\eqref{eq:R0} relates $R_0$ to the collective
decay rate $\gamma_\sjE$, which then implies
\begin{equation}
  \label{eq:1}
  \lim_{N\rightarrow \infty} \gamma_{\sjE} =
  \frac{3}{2\pi(a/\lambda)^2} \Gamma_{\sjE} \textrm{.}
\end{equation}
This precisely corresponds to the asymptotic value of $\gamma_{\sjE}$
for large array sizes shown in Fig.~\ref{fig:gamDelt_v_N}.

The radiative decay rate $\gamma_{\sjM}$ of the PMM mode, on the other hand, asymptotically
approaches zero.
Since the magnetic dipoles are perpendicular to the array, they do not
emit in the forward and backward directions, but as the array size
increases, interference of radiation from the various meta-atoms  diminishes PMM emission in other directions.
The PMM mode therefore has zero emission in an infinite array.
This finding is consistent with Fig.~\ref{fig:gamDelt_v_N} which shows
that asymptotically $\gamma_{\sjM}\propto 1/N$, and the value of $\gamma_{\sjM}$ can also be extrapolated for large arrays.

\section{Narrowing of the transmission resonance}
\label{sec:decr-linew-incr}

In this appendix, we show how increasing the size of the metamaterial
array allows one to narrow the spectral width of the transmission
window and increase the group delay of a pulse passing through the
array. For simplicity, we assume $\delta_{\sjM} - \delta_{\sjE} \ll
\gamma_{\sjE}$ so that we can neglect any difference between the PME and
PMM resonance frequencies.
We further assume that the array is sufficiently large that
$\gamma_{\sjE}$ can be approximated by its asymptotic value
[Eq.~\eqref{eq:1}] so that $R_0 \approx -1$.
In doing so, one finds that a local maximum in transmittance occurs on
PMM resonance.
The reflectance amplitude on PMM resonance is thus,
\begin{equation}
  \label{eq:R_delta_M}
  R(\delta_{\sjM}) \approx - \frac{\gamma_\sjE
    \gamma_{\sjM}/4}{(\delta\omega)^2 - \gamma_{\sjE} \gamma_{\sjM}/4}
\end{equation}
When the asymmetry of ASRs satisfies $(\delta\omega)^2 \gg \gamma_{\sjE}\gamma_{\sjM}$, reflectance on
PMM resonance is suppressed, and transmittance is enhanced.

To determine the properties of the transmission window, we assume
the asymmetry is large enough so that one can express the transmittance properties to zeroth order in
$\gamma_{\sjE}\gamma_{\sjM} / (\delta\omega)^2$.
Expanding $T$ in $\Delta-\delta_{\sjM}$ one finds
\begin{equation}
  \label{eq:T_Taylor}
  T(\Delta) \approx 1 + \frac{i}{2}\frac{\gamma_{\sjE}\left(\Delta - \delta_\sjM\right)}{(\delta\omega)^2} -
  \frac{1}{4}
  \left(\frac{\gamma_{\sjE}\left(\Delta -
        \delta_\sjM\right)}{(\delta\omega)^2}\right)^2 \textrm{ .}
\end{equation}
The width of the transmission window is determined by the intensity
transmittance $|T|^2$, which can be approximated near PMM resonance
using Eq.~\eqref{eq:T_Taylor}.
From Eq.~\eqref{eq:w_def}, one finds the approximate resonance width
\begin{equation}
  \label{eq:sigma}
  w = 4\sqrt{\log 2}\frac{(\delta\omega)^2}{\gamma_{\sjE}}
  \textrm{ .}
\end{equation}
\vspace{0.01em}\newline
The quality, or inverse spectral width, of the transmission resonance
therefore varies as $1/(\delta\omega)^2$.
Similarly, the group delay is approximated by [Eq.~\eqref{eq:tau_g}]
\begin{equation}
  \tau_g =  \frac{\gamma_{\sjE}}{2(\delta\omega)^2}
\end{equation}
The group delay thus also scales with $1/(\delta\omega)^2$.

Since both the quality of the transmission window and the group delay
of a resonant pulse scale inversely with $(\delta\omega)^2$, one could
increase both quality and group delay by reducing the asymmetry.
If $\delta\omega$ becomes too small, however,  the intensity transmittance on PMM
resonance decreases.
To lowest order in $\gamma_\sjE\gamma_\sjM/(\delta\omega)^2$,
deviation of the peak transmittance from 1, $\xi \equiv 1 -
|T(\delta_\sjM|^2$, is
\begin{equation}
  \label{eq:T2Dev}
  \xi \approx \frac{1}{2} \frac{\gamma_\sjE
    \gamma_\sjM}{(\delta\omega)^2} \textrm{.}
\end{equation}
Thus, even as a smaller $\delta\omega$ enhances the resonance quality
and group delay of a resonant pulse, it reduces the peak transmittance.
To ensure the peak transmittance remains sufficiently high, we define
a maximum tolerable deviation $\xi_{\mathrm{max}} \ll 1$ such that $\gamma_\sjE
\gamma_\sjM/[2(\delta\omega)^2] < \xi_{\mathrm{max}}$.
This imposes a lower bound on the degree to which $\delta\omega$ can
be reduced while still maintaining the transmission resonance
\begin{equation}
  \label{eq:delta_omega_lower}
  (\delta\omega_{\rm min})^2 >
  \frac{\gamma_{\sjE}\gamma_{\sjM}}{2\xi_{\mathrm{max}}} \textrm{.}
\end{equation}
But, as Fig.~\ref{fig:gamDelt_v_N} illustrates, $\gamma_{\sjM}\propto 1/N$, decreasing as the array gets
larger. One can therefore decrease the lower bound on $(\delta\omega)^2$ by
increasing $N$.
For example, to quadruple the resonance quality and group delay while
maintaining the peak transmittance, one would simultaneously quadruple
the number of ASRs (doubling the side lengths of the array) and halve
the asymmetry $\delta\omega$.

%

\end{document}